\documentstyle[12pt,epsfig]{ioplppt}
\begin{document}
\jl{4}
\title{Characteristics of air showers produced by extremely high energy 
gamma-rays}[Characteristics of EHE gamma-ray air showers]
\author{A V Plyasheshnikov\dag  and 
F A Aharonian\ddag}

\address{\dag Altai State University, Barnaul, Russia}

\address{\ddag\ Max-Planck-Institut f\"ur Kernphysik, 
Heidelberg, Germany}

\begin{abstract}
The  technique of adjoint cascade equations has been applied to 
calculate  the properties of air showers produced by extremely 
high energy  (EHE) $\gamma$-rays in the energy range 
$10^{18}-10^{22}$~eV.  The high intrinsic accuracy,  
combined with very modest  (compared with the  
traditional Monte Carlo codes)   computational time 
requirements,  make this method as an effective 
tool for detailed study of development of   EHE showers 
in the Earth's atmosphere.  In this paper  a wide range of 
parameters of $\gamma$-ray  induced  showers  are analysed
taking into account  two independent effects which
become crucial for the cascade development
in the EHE regime  --   the  Landau-Pomeranchuk-Migdal (LPM) 
effect and  the interaction of primary $\gamma$-rays  with 
the geomagnetic field (GMF).  Although the LPM effect leads  to 
dramatic modifications  of  shower characteristics, especially
at primary energies exceeding   $10^{19} \ \rm eV$,
the GMF  effect, which  starts to ``work''   approximately at same
energies,  prevents, to a large extent,  the LPM effect by converting
the primary  $\gamma$-ray  into a bunch of  
synchrotron $\gamma$-ray photons with energies 
effectively below  the  threshold of the LPM effect.  
This bunch of the secondary photons  
hits the atmosphere  and creates a large number of 
simultaneous showers.  The superposition of these independent 
showers mimics a single  shower 
with energy $E=\Sigma E_{\rm i} \simeq E_0$, but without
the signatures of  the LPM effect.   This makes the longitudinal 
profile of such a   composite electromagnetic shower quite 
similar to the longitudinal  profile  of hadronic showers.
At the same time, the number of 
muons  as well as their lateral distribution 
differ  significantly from the corresponding parameters of   
proton-induced showers.   In the ``gamma-ray bunch''  
regime, the total number of muons is less, by a factor of 
5 to 10,  than the number of muons in hadronic showers.
Also,   compared with the hadronic showers, 
the electromagnetic showers are characterized  
by  a significantly  narrower  lateral distribution  of muons.
Even so,  for inclined EHE $\gamma$-ray showers 
the density of muon flux at large  distances from the shower core 
($\ge 1000$~m) can  exceed the electron  density.

\end{abstract}

\maketitle

\section{Introduction}
The question of the origin of most energetic particles 
observed in cosmic rays at energies $E\ge 10^{18}$~eV, often called 
Extremely High Energy cosmic rays (EHECRs), is a subject of intensive 
astrophysical speculations and debates. Within the conventional 
acceleration (or "bottom-up")  scenarios,  the powerful radiogalaxies, 
active galactic nuclei and clusters of galaxies (see e.g.
\cite{cite1,cite2,rachen}),    as well as transient events like the sources
of $\gamma$-ray bursts \cite{cite3,cite4,cite5} are believed to be the 
most probable sites of production of EHE cosmic rays. These models have, 
however, a difficulty connected with the large,  typically 
$\geq 100 \ \rm Mpc$  distances to the most prominent 
representatives of these  source populations. Indeed, if the observed 
EHE cosmic rays are protons produced in distant extragalactic sources, 
their interaction with the 2.7 K cosmic microwave background 
radiation (MBR) should give rise to the so-called Greisen-Zatsepin-Kuzmin 
(GZK) spectral cutoff.  Depending on the characteristic distance 
scale to the ensemble of sources responsible  for the bulk
of the observed EHE cosmic ray  flux, the position of the cutoff is expected
between $E\sim 5 \times 10^{19}$ and $10^{20}$~eV (see e.g. 
Refs.\cite{cite6,cite7,stanev1}). Therefore, the registration  of the  
EHECR events close to $10^{20}$~eV 
by  Haverah Park,  Fly's Eye, Yakutsk,  and AGASA 
detectors  (for review see 
Refs.\cite{cronin,watson})   
pose a serious challenge for any of these 
conventional models. This difficulty initiated  
a quite different approach to the solution of the problem of origin of 
EHE cosmic rays -- the hypothesis of {\it non-acceleration} (or "top-down") 
scenario which assumes that the highest energy cosmic rays are result 
of decay of hypothetical massive relic particles originating from 
early cosmological epochs (see e.g. Ref.\cite{cite10} and references
therein). A distinct feature of the "top-down" models of EHECRs is an 
unusually  high content of $\gamma$-rays with the 
gamma/proton  ratio  $\ge 1$ at $E\ge 10^{20}$~eV \cite{cite11,cite12,cite13}.
Thus the gamma/proton ratio can serve as a conclusive diagnostic 
tool for observational proof (or rejection) of the "top-down"
scenarios.

Actually, a non-negligible content of $\gamma$-rays into the EHE
cosmic rays 
is expected also in the conventional particle acceleration models. 
Within these models the highest energy $\gamma$-rays are contributed 
through  the decay of secondary $\pi^0$-mesons produced 
at interactions of  EHECRs with 2.7 K MBR 
\cite{Wolfendale,Aharonian1}. If the average magnetic 
field in the intergalactic medium is less than $10^{-12}$~G, the 
development of the ultrarelativistic electron-photon cascades in
the field of 2.7 K  MBR makes the attenuation length of EHE 
$\gamma$-rays at $E\ge 10^{20}$~eV  larger than the 
attenuation length of protons. Therefore, the $\gamma$-ray content
in EHE cosmic ray  flux depends strongly  on the typical distances  
to the cosmic ray  sources, and on the strength of  
the intergalactic magnetic field. 
It could be as large as 10 per cent in the 
case of homogeneous distribution of the EHECR sources in the 
Universe, but less than 1 per cent if the bulk of the observed 
EHECR flux is contributed by relatively nearby sources  concentrated
in our Local Supercluster  of Galaxies. Thus the 
measurements of the gamma/proton ratio in EHE cosmic rays
by forthcoming  powerful instruments like the Auger Observatory or the 
Telescope Array  (see e.g. Ref.\cite{cronin,watson}) would not only 
allow crucial tests
to distinguish between the "top-down" and "bottom-up" scenarios,
but also can provide important constraints on the spatial 
distribution of the EHECR accelerators within the conventional
"bottom-up" models.
       
The effective separation  of $\gamma$-ray induced showers
from the proton-induced  showers is possible because of 
noticeable differences 
in the development of electromagnetic and hadronic cascades.
Starting from energies of about 10~TeV, the low-energy muon 
content becomes  a reliable criterion for recognition of  electromagnetic 
showers  \cite{Stanev, Edwards, Hillas, Procureur}).
At energies $E\ge 10^{16}$~eV  the content of muons in the 
electromagnetic showers increases with an increase of energy 
significantly faster than in the hadronic cascades \cite{Aharonian3}.
The results of this paper based on detailed numerical 
calculations of characteristics of electromagnetic showers,
using the so-called adjoint equation technique 
(see for review \cite{Uchaikin}),
generally confirm this effect. However, the present detailed 
numerical study shows that the prediction of  Ref.\cite{Aharonian3}
concerning the absolute muon number, based on an approximate 
analytical approach,  was significantly   overestimated.    
 
At primary energies above $10^{19}$~eV the 
Landau-Pomeranchuk-Migdal (LPM) effect \cite{cite20,cite20b}   
leads to  significant  suppressions of the Bethe-Heitler 
cross-sections for the  pair-production  and 
bremsstrahlung processes in the Earth's atmosphere, 
and thus  dramatically changes the character of development 
of electromagnetic  showers. In practice, however, the interesting 
features of electromagnetic showers caused 
by the LPM effect do not appear  in a "pure" form, but
are largely  compensated by another effect 
connected with interactions of EHE $\gamma$-rays with 
the geomagnetic field (GMF) \cite{Mcbreen}.  Remarkably, for the 
combination of relevant  parameters, namely the strength and 
orientation of the geomagnetic field from one side, and the 
height, density and the mean atomic number  of the Earth's 
atmosphere from another side, these two effects start to   
"operate" almost simultaneously,  at  energies above several 
times $10^{19}$~eV.  The interaction of $\gamma$-rays
with the GMF by $e^+, e^-$ pair 
production  does not
imply, however,  absorption of primary photons. In fact, the 
secondary (pair produced) electrons and positrons in the same 
geomagnetic field quickly produce a bunch  of synchrotron
photons with energies typically between 
$10^{15}$ and $10^{19}$~eV.   
The secondary photons initiate a large number of simultaneous    
electromagnetic cascades in the 
atmosphere. Because almost the whole energy 
of the primary photon is re-distributed between the 
synchrotron  photons, the superposition of these showers mimics  
a single  electromagnetic shower with total energy rather close 
to the energy of the primary particle.   
Moreover, because of the relatively low energies of 
the bunch  photons, the LPM  effect is essentially "switched off", 
and the longitudinal profile of the shower returns to the  
``standard''  shape of electromagnetic showers 
without the signature  of the LPM effect \cite{Aharonian3}.
For typical strengths of the GMF of about 
$0.2-0.6$~G and for energies of primary $\gamma$-rays 
$E\ge 10^{20}$~eV, this interesting interference of two strong 
effects becomes unavoidable almost at all 
arrival directions of primary $\gamma$-rays. Therefore,
any study of electromagnetic showers of such high energies requires  
thorough treatment of both the LPM and GMF effects.
The depth profiles of the electronic component of EHE electromagnetic 
showers including these two effects have been studied by several 
authors  \cite{Aharonian3,Goncharov,Halzen,cite24,Jussieu}.
Recently the characteristics of the muon component  have been 
calculated by  Capdevielle  et.al. \cite{Capdevielle} using 
the CORSIKA Monte Carlo code. They took into account the   
LPM effect, but ignored the interactions of the primary 
$\gamma$-rays with the geomagnetic field.  
 We generally confirm the results of 
these studies,  but significantly extend the range of calculated
characteristics. In particular, we calculated  such basic 
parameters  of electromagnetic showers as  the electron size,
the depth profiles of cascade particles, 
the  content and the lateral distribution of muons  and compared 
them  with the relevant parameters of hadronic showers.   
 
\section{Interaction processes}
In calculations of  characteristics of electromagnetic air showers 
we take into account the following  interaction  processes -- 
the bremsstrahlung 
and the ionization  losses for electrons and positrons; the pair production
and the Compton scattering for photons. The  LPM effect 
\cite{cite20,cite20b} is 
taken into account for the bremsstrahlung    and the pair production processes.
In addition, we take into consideration the interaction 
of EHE $\gamma$-rays  with the geomagnetic field. 
The  cross-sections of magnetic cascade processes 
(the pair production and the  synchrotron radiation) are calculated 
in accordance with \cite{Angelov, Akhiezer}.

We calculate the spatial distribution of the GMF  within the 
framework  of the dipolar central model (DCM;  see  e.g.  \cite{Chapman}). 
In table~1  we present the  GMF for two 
proposed sites  of the Auger Observatory
in Northern and Southern Hemispheres \cite{Auger}. 
Besides the DCM results 
we present in table~1 the strengths of GMF  obtained in Ref.\cite{Cillis} 
by use of the data of the International Geomagnetic 
Reference Field \cite{IGRF}.
One can see that the dipolar central model  
overestimates significantly  the $B$-field at the Southern Hemisphere 
site. Therefore in  calculations we multiply the  
DCM results on $B$ by special correction factors
providing the coincidence  with  data of
\cite{Cillis} near the Earth surface. With such a correction
the field strength  for the Northern Hemisphere site of the Auger
Observatory  is larger by a factor of  $\simeq 2$ compared with 
the Southern Hemisphere site (table~1). 

\begin{table}
\caption{The strength of the geomagnetic field  (in Gauss) 
near the Earth surface}
\begin{indented}\item
\begin{tabular}{@{}lcc} \br
Site location        &  El Nuhuil, &  Millard County,   \\
                     &Mendoza (Argentina)& Utah (USA) \\ \mr
Geographical coordinates &     35.2$^o$~S, 69.2$^o$~W&
39.1$^o$~N, 112.6$^o$~W   \\ 
Observation level, g/cm$^2$ &   890&  870  \\ 

DCM data                             &     0.376&   0.497  \\ 
Data of   \cite{Cillis}            &     0.250&   0.528  \\ \br
\end{tabular}
\end{indented}
\end{table}

In figure~1 we show  examples of the spatial profile of the GMF component 
($B_{\rm n}$)  perpendicular  to the arrival direction of the air shower 
(note that this component of the field is the only parameter which 
is needed to calculate the cross-sections of both 
the pair-production and synchrotron processes). The 
parameter  $t$ in figure~1  is the distance between the Earth surface
and the observation point measured along the air shower  
arrival direction \footnote{Note that for inclined showers 
the B-field depends also on the azimuth angle. However, 
for simplicity,   for such showers we assume  
a fixed azimuth angle corresponding to the  
arrival  direction from the nearest geomagnetic pole.}.
One can see that $B_n$ increases with the zenith angle $\theta$ and 
decreases rapidly as a function of distance $t$.

For  calculations of basic characteristics of 
photonuclear interactions of $\gamma$-rays
we  performed special simulations with 
use of   the SOPHIA generator \cite{Muecke}. 
This generator provides  a comprehensive tool  for simulations of 
photohadronic processes. In particular,  this code includes
the excitation  and decay of baryon resonances, the direct production
of pions and the multiple production of hadrons.  
To simplify the calculation technique we assumed that all unstable 
particles generated in the photoproduction  process (except for 
muons) immediately decay at the point of their generation,
i.e. we neglect the finiteness of life time of such particles
\footnote{Our calculations  show that this is a quite 
acceptable  assumption, if one deals with muon detectors with 
sufficiently small threshold energy,   $\le 1$~GeV.}.
In this approximation all secondary muons are assumed to be emitted at 
the  point of the primary $\gamma$-ray interaction. Thus,  the inclusive 
differential (over the muon kinetic energy ($T$) and the muon emission 
angle  ($\vartheta$)) cross-section $w_{\mu}(E_{\gamma},T,\vartheta)$ 
becomes  the only input parameter. 
This cross-section is normalized  as: 
\begin{equation}
   2\pi\int d\vartheta\sin\vartheta \int 
   w_{\mu}(E_{\gamma},T,\vartheta) dT = \sigma_{\rm ph}\cdot\bar{n}_{\mu} 
\end{equation}
where $\sigma_{ph}$ is the total cross-section of photoproduction, 
$\bar{n}_{\mu}$ -- the mean multiplicity of muons. 
Following to recommendations of \cite{Stanev}
we define the quantity $\sigma_{\rm ph}$ as 
$\sigma_{\rm ph}= \sigma_{\gamma \rm p}
\cdot  A ^{0.91}$ where $\sigma_{\gamma \rm p}$ is the total cross-section of 
$\gamma$-ray - proton interaction (we use for this quantity data from
\cite{Muecke}), $A$ is the mean atomic number of air nuclei.

Besides the  electromagnetic cascades we have calculated 
also the characteristics of  proton-induced showers.
A standard quark-gluon-string (QGS) model is applied in this case 
to describe  the hadronic interactions \cite{Shabelsky1,Shabelsky2}.

\section{Technique of adjoint cascade equations}

The calculations  presented below are obtained  with
the help of numerical solution of the adjoint cascade 
equations. In this section we describe shortly the main features 
of these equations\footnote{More detailed description of the
adjoint equation technique is presented in  \protect\cite{Uchaikin}.}.  

Let us consider as an example the  system of 
adjoint equations that describe  the longitudinal 
development of the electromagnetic cascade. Studying the 
longitudinal development  of such a cascade one can neglect 
the lateral displacement of secondary particles and their
angular distribution induced mainly by the multiple 
Coulomb scattering.
The equation system has in this case   the following form
\begin{equation}
\label{eq1}
\lambda_e\frac{\partial f}{\partial t} + f - 
\int_{E_0}^{E} w_{ee}(E,E') f(t,E')dE' -
\int_{E_0}^{E} w_{e\gamma}(E,E')g(t,E')dE' = \lambda_e F,
\end{equation}

\begin{equation}
\label{eq2}
\lambda_{\gamma}\frac{\partial g}{\partial t} + g - 
\int_{E_0}^{E} w_{\gamma e}(E,E') f(t,E')dE' -
\int_{E_0}^{E} w_{\gamma\gamma}(E,E')g(t,E')dE' = \lambda_{\gamma} G.
\end{equation}

The adjoint functions $f$ and $g$ in  Eqs. (\ref{eq1}), and (\ref{eq2})
describe  the contributions (averaged over random realizations of the 
cascade development)  to the response of some detector from a 
cascade generated by a  primary  electron ($f$) or a photon ($g$) 
of energy $E$; $t$ is the distance between 
the observation level and a point of the primary particle 
appearance (measured along the particle arrival direction); 
$E_0$ is the particle detector threshold energy; $\lambda_{\alpha}$ 
is the mean free pathlength for the cascade particle of type 
$\alpha$ ($\alpha=e \ \rm or \ \gamma)$; 
functions $w_{\alpha\beta}(E,E ')$  
define the differential 
(over the energy $E '$) spectra of secondary particles 
of type $\beta$ ($\beta = e \ \rm or \ \gamma$).

Properties of  the detector are defined in Eqs.~ (\ref{eq1}) 
and (\ref{eq2}) by 
the right  hand side functions  $F$ and $G$  and the 
boundary conditions    $f(t=0,E)$ and $g(t=0,E)$. For example, 
if the detector measures at the observation level the total number 
of cascade  electrons with energy  $E \ge E_0$, then
\begin{equation}
\label{eq3}
F(t,E)\equiv G(t,E)\equiv 0; \quad  
f(t=0,E)=H(E-E_0), \quad g(t=0,E)\equiv 0
\end{equation}   
where   
$H(x)\equiv 1$ for $\quad  x\ge 0$  and  
$H(x)\equiv 0$ for $\quad x<0$. 

In the case of calculation of characteristics of muons
produced in the electromagnetic cascade due to nuclear 
interactions of cascade photons one can use a 
system of adjoint equations  similar to Eqs.~(\ref{eq1}) and 
(\ref{eq2}), 
but having the different form of right hand side terms  
and boundary conditions. For example, if the detector 
registers  the  muons transversing a circle of a given 
radius ($R_0$) with the centre at the shower axis and 
orientated perpendicularly to  this axis,  we have 
\begin{equation}
   f(t=0,E)=g(t=0,E)=0, \qquad F(t,E)=0,
\end{equation}
\begin{equation}
\label{eq55}
G(t,E)=2\pi n_0\int d\vartheta\sin\vartheta\int
w_{\mu}(E,T,\vartheta)p_{\rm d}(T,\vartheta) 
p_{\rm m}(T,\vartheta)
H(\tilde{T}-T_{\rm th}) dT \ ,
\end{equation}
where $w_{\mu}$ is the differential  cross-section 
of muon photoproduction defined in section~2,  
$T_{\rm th}=E_0 - m_\mu c^2$ -- the detector threshold 
kinetic energy, $n_0$ -- 
the concentration of air nuclei. The function  
\begin{equation}
p_{\rm d}(T)=\exp\{-\cos\vartheta ^{-1} \int_{\tilde{T}}^{T}
[\lambda_{\rm dec}(T')\cdot \epsilon(T')]^{-1} dT' \}
\end{equation}
defines the probability for the muon to survive on the way from
its production  point to the observation level. $\tilde{T}$ is 
the energy  of the muon at the observation level found from 
the following equation 
\begin{equation}
t/\cos\vartheta=\int_{\tilde{T}}^{T} dT'/\epsilon(T') \ , 
\end{equation}
$\epsilon$ is the ionization  loss rate, and 
$\lambda_{\rm dec} $ is the decay mean free
pathlength.

The function $p_{\rm m}$ is the probability  
of the muon to escape  from the 
circle due to the multiple (Coulomb) scattering.
We calculate this function  using  the expression  
for the probability  distribution  function  of muon lateral displacements 
based on the solution of   a special kind of the adjoint transport equation
obtained by  \cite{Kolchuzhkin}.   

For solution of adjoint cascade equations we use a numerical
method which is described in 
\cite{Plyasheshnikov}. This method  has been applied to solve 
a wide range of problems connected with  cosmic ray physics 
(see for review  \cite{Uchaikin}). 
The method \cite{Plyasheshnikov} provides an 
accuracy of solution of adjoint equations better than 
a few per cent and gives  results for an  arbitrary region 
$(E_0,E_{\rm max})$ of the primary energy in  
{\it one run}  of calculations.   
The computational time consumed 
by the method  is quite small and depends   weakly (logarithmically) 
on the primary energy.  For example, to calculate the number  of 
cascade electrons (or photons) for a fixed altitude of the observation 
level and the  primary energy range $(10^9,10^{22})$~eV
one needs    only a few minutes of the 
1 GHz IBM PC computational time. 

The  results for proton-induced air showers 
also have been obtained by  solution of adjoint cascade equations 
with  the help of the numerical method  
developed in  Ref. \cite{Plyasheshnikov}.  Here we do not 
present  the expressions for these equations, because they are 
quite  similar to Eqs. (\ref{eq1}) and (\ref{eq2}).  They can be found  
in \cite{Uchaikin}.

\section{Comparison with previous works}
A number of  comparisons of air shower characteristics 
with  similar results of previous studies  has been performed to  
test the  computational code.    
In figure~2 the cascade curves of photons are presented for the 
$\gamma$-ray induced showers developing   in the homogeneous magnetic 
field (without presence of matter). One can see  
a perfect  agreement with calculations
of A.Goncharov \cite{Goncharov}.

Figure~3 represents the cascade curves of electrons 
for air showers initiated by 
primary $\gamma$-rays of energy $10^{20}$~eV,  taking into account 
interactions with the geomagnetic field and the LPM  effect. 
Three different combinations of the GMF and LPM effects 
are shown by curves 1 (no LPM, no GMF),  2 (only LPM),
3 (both  LPM and  GMF).  A  quite satisfactory   
agreement   with the results  of   
Ref. \cite{Aharonian3} is seen for all 3 cases.

In table~2 we present  the number of  muons  
at the sea level versus  the total number of electrons 
for air showers created by primary $\gamma$-rays with energy 
below  $10^{16}$~eV ($N_{\rm e} \leq 10^7$). Our results  are compared 
with similar results of Edwards et.al. \cite{Edwards}.    
Note that the calculations of Ref. \cite{Edwards}, 
as well as many other studies
in this energy region, were  performed  using  simplified 
treatments  of photohadronic interactions.  Namely, it was assumed that the
$\gamma$-hadron interaction is identical to the    
inelastic interaction of the pion of the same energy.  Nevertheless, 
a quite satisfactory agreement is seen between results 
presented in the table.

\begin{table}
\caption{The number of muons with energy $T\ge $ 1~GeV as a 
function of the total number of electrons at the sea level 
for air showers  created by inclined $\gamma$-rays with  
zenith angle $25^{\circ}$}.
\begin{indented}\item
\begin{tabular}{@{}lccccc} \br
$N_e$   &
                       $10^5$&   $3\cdot 10^5$&          $10^6$&
                $3\cdot 10^6$&           $10^7$ \\ \mr
$N_{\mu}$, present study&       
             $0.29\cdot 10^3$&$0.77\cdot 10^3$&$0.20\cdot 10^4$&
             $0.51\cdot 10^4$&$0.15\cdot 10^5$    \\   
$N_{\mu}$, \cite{Edwards}&     
             $0.33\cdot 10^3$&$0.92\cdot 10^3$&$0.27\cdot 10^4$&
             $0.65\cdot 10^4$&$0.16\cdot 10^5$     \\ \br
\end{tabular}
\end{indented}
\end{table}

In table~3 we present the calculation results  for the number of 
photoproduced muons
with kinetic energy $T\ge 0$.  For comparison we show also 
the results of calculations performed by us using the CORSIKA 
code \cite{Heck}.  It should be noted that the results in table~3 
correspond to a limited primary energy region which allows us 
to use the CORSIKA option 
based on the ``complete'' Monte Carlo  method, i.e.  without  the 
application of the ``thinning technique'' \cite{Hillas2}.   
It is seen that discrepancy of results
presented in the table does not exceed  20 \%.  

The  ``complete'' Monte Carlo  method requires a very large 
computational time which does not allow extension of simulation to 
extremely high energy region.  To reduce the computational time one has 
to  apply the  ``thinning technique'' \cite{Hillas2}.  Such calculations 
were performed by Capdevielle et al.   \cite{Capdevielle} using the 
same  CORSIKA simulation code for EHE primary $\gamma$-rays.
In table~4 we compare our results on the muon content in EHE $\gamma$-ray 
induced showers  with similar results  of Ref. \cite{Capdevielle}.
Note that Capdevielle et al.   \cite{Capdevielle} included in 
their calculations the  LPM effect but ignored the effect of 
interactions of $\gamma$-rays with 
the geomagnetic field. Therefore in table~4  we show our results  
without the   
GMF effect as well.  The results are in good agreement within 10 \%.   

\begin{table}
\caption{The muon content in air showers created by primary
$\gamma$-rays for $z_{\rm obs}=$890 g/cm$^2$. Results of the numerical 
solution of adjoint equations are compared with simulation results
obtained by the CORSIKA code.}
\begin{indented}\item
\begin{tabular}{@{}lcccccc} \br
Primary energy of air shower, eV  
& $3\cdot 10^{13}$  &   $10^{14}$   & $3\cdot 10^{14}$& 
$10^{14}$ & $3\cdot 10^{14}$   \\
Zenith angle of arriving $\gamma$-ray, degree&  
0  &     0   &     0&  45 &   45   \\ \mr
$N_{\mu}$, solution of adjoint equations& 
13.9&   56.4  &   199&  21.9&  69.8  \\
$N_{\mu}$, simulations with CORSIKA    & 
11.6&   50.2  &   174&  26.1&  79.0  \\ \br

\end{tabular}
\end{indented}
\end{table}

In figure~4 we  present the radial distributions of the photoproduced 
muons for vertical air showers created by $\gamma$-rays
with energy $10^{15}$~eV.  Results of the numerical solution
of adjoint equations are compared with the simulations made 
by  the ``complete'' Monte Carlo method  using the 
CORSIKA code. \footnote{The    ``thinning technique'' 
\cite{Hillas2} does not  allow  adequate calculations of the radial
distributions  of muons
in EHE electromagnetic air showers (see section~6), therefore 
the comparison  presented in the table is limited by the 
primary energy  $10^{15}$~eV.}. 
A perfect agreement of the results is achieved.

\begin{table}
\caption{The number of muons with energy $T\ge 0$ as a 
function of energy of the primary $\gamma$-ray 
for inclined ($45^o$) air showers.
The LPM effect is included, the geomagnetic field is disabled.}
\begin{indented}\item
\begin{tabular}{@{}lcccc} \br
$E_{\gamma}$, eV   &
                       $10^{17}$&           $10^{18}$&          $10^{19}$&
                    $10^{20}$  \\ \mr
$N_{\mu}$, present study&       
             $4.5\cdot 10^4$&$6.1\cdot 10^5$&$8.2\cdot 10^6$&
             $7.5\cdot 10^7$     \\   
$N_{\mu}$, \cite{Capdevielle}&     
             $5.0\cdot 10^4$&$6.6\cdot 10^5$&$8.1\cdot 10^6$&
             $7.9\cdot 10^7$       \\ \br
\end{tabular}
\end{indented}
\end{table}

In figure~5 we present the electron cascade curves   for proton 
induced air showers.  It is seen that  our results are 
in agreement within 20 \% with simulations performed 
in the framework of the QGS model  \cite{Lagutin}.

\begin{table}
\caption{The number of muons with energy $T\ge 0$ 
for vertical air showers
created by primary protons. $Z_{obs}=$920 g/cm$^2$. }
\begin{indented}\item
\begin{tabular}{@{}lccccc} \br
$E_p, eV$   &       $10^{18}$&$3\cdot 10^{18}$&
                    $10^{19}$&$3\cdot 10^{19}$&       $10^{20}$ 
\\ \mr
$N_{\mu}$, present study&
             $0.71\cdot 10^7$&$0.19\cdot 10^8$&
             $0.56\cdot 10^8$&$0.15\cdot 10^9$&$0.43\cdot 10^9$ 
\\  
$N_{\mu}$, \cite{Nagano}&     
             $0.68\cdot 10^7$&$0.20\cdot 10^8$&
             $0.51\cdot 10^8$&$0.16\cdot 10^9$&$0.45\cdot 10^9$
\\ \br
\end{tabular}
\end{indented}
\end{table}

In table~5  we present the muon content in air  showers 
initiated by vertical protons.  Our results on this quantity are compared 
with similar results of  Nagano  et.al. \cite{Nagano} obtained   
by the CORSIKA simulation  code with use of the QGSJET 
generator of  hadron interactions.  The difference between these 
two methods does not exceed 10 \%. 

\section{Longitudinal development of air showers}
\subsection{Muon component of $\gamma$-induced air showers}
   
In the energy region below $\simeq 10^{16}$~eV  
the muon content in electromagnetic air showers has been 
studied by many authors (see e.g.
\cite{Stanev,Edwards,Hillas,Procureur,Danilova,Chatelet}).
However, the calculations , performed in this energy region, 
are based, as a rule,  on  rather simplified treatments of the 
photoproduction process. For example, it was often assumed that 
the $\gamma$-hadron  interactions were  identical to the  
inelastic pion interactions.  Recently the photohadronic interactions 
have  been implemented into the simulation codes CORSIKA \cite{Heck}
and AIRES \cite{Sciutto2} allowing 
accurate calculations of characteristics of muons  in EHE 
electromagnetic air showers  \cite{Capdevielle,Cillis,Ave}.  
However, the interaction with
the geomagnetic field has not yet been included in these codes.
The latter reduces essentially the applicability of
calculation results on the content of photoproduced muons  
calculated with the help of these codes.  In the present study
we calculate the shower characteristics using  the 
technique of adjoint cascade equations and considering interactions of 
 primary $\gamma$-rays with a composite target  consisting of two 
parts - the   geomagnetic field and the Earth's atmosphere.  

In  figure~6 we present the dependences of the total number 
of muons ($N_{\mu}$) at the observation level 
on the energy of primary  $\gamma$-ray for vertical and inclined  
showers and for different assumptions  about 
the LPM effect  and the geomagnetic  field. The 
following conclusions can be obtained from this figure.
\begin{itemize}
\item
If one  uses  the Bethe-Heitler's  cross-sections for 
high energy interactions of cascade particles, 
the energy dependence  of $N_{\mu}$ 
is close to the power law, i.e.  $N_{\mu}\sim E_{\gamma}^{\beta}$
with $\beta \simeq 0.9-1.0$.  
\item
The LPM  effect changes dramatically the  
energy dependence of $N_{\mu}$. 
The growth of  $N_{\mu}(E_{\gamma})$ becomes 
less rapid; moreover a local maximum  arises near $E_{\gamma}\simeq 
10^{20}-10^{21}$~eV, after which $N_{\mu}$ decreases with $E_{\gamma}$
(curves ``2'' in figures~6a and 6b).
\item 
The GMF (without the LPM effect)  
does not essentially change  the energy dependence of  
$N_{\mu}(E_{\gamma})$. It remains to be continuously 
growing for both  the vertical and inclined showers.
\item
After including of the interactions with the geomagnetic field, 
the LPM effect does not affect considerably the  
$N_{\mu}(E_{\gamma})$ dependence. However, some deviation
from the power law takes place in the energy region 
$10^{19}\div 10^{21}$~eV. This deviation is connected with the fact,
the LPM effects becomes effective  a bit earlier (i.e. at lower 
energies) compared with the  GMF effect. 
\item   
Behaviour  of  $N_{\mu}(E_\gamma)$ for 
inclined air showers is generally similar to the one of the vertical showers.
However, a local maximum of $N_{\mu}(E_{\gamma})$ 
dependence, connected with the LPM effect,  is observed for
larger  primary energies.   
Besides, for inclined showers 
the simultaneous influence of the LPM effect 
and the GMF  can lead to a slight reduction of 
$N_{\mu}$ in the high energy region,  whereas  for vertical 
air showers these effects  lead to  a  growth 
of $N_{\mu}$. 
\end{itemize} 

The GMF influence on $N_{\mu}$ as well as on other shower 
characteristics has a simple explanation.   
For sufficiently high product $E_\gamma \cdot B_{\rm n}$ 
the primary $\gamma$-ray cannot freely enter the Earth's atmosphere
due to the pair production in the geomagnetic field. The secondary 
(pair-produced) electrons 
and positrons immediately interact with the same 
field through the synchrotron radiation.
The free pathlength of these electrons is  rather small 
(see figure~7) compared with the scale of the magnetosphere,
and  weakly depends on the particle  
energy.  Thus, these electrons   emit on their 
way to the Earth a large number of relatively 
low energy photons  (according to figure~7 -- one photon per 
every 10-100~km of 
the trajectory).  On the other hand,  the mean free pathlength 
for the pair production grows very 
rapidly with reduction of the photon energy (see figure~7). 
As a result, the energy of primary  $\gamma$-ray is distributed  
among a large number of 
secondary photons with relatively small energies not 
able to interact effectively with the geomagnetic field
(this effect is illustrated by table~6). Thus, instead of a single  
$\gamma$-ray photon we have at the upper border  of the atmosphere 
a 'bunch'  of low energy photons ($E \leq 10^{19}$~eV) 
which thus appears below the  effective threshold of the LPM effect.   
    
\begin{table}
\caption{The mean number of cascade synchrotron photons 
produced in GMF by a vertical 
primary $\gamma$-ray of energy $10^{21} \, \rm eV$,  
and  entering the  Earth's  atmosphere
with energy greater  than $E_{\rm th}$.}
  
\begin{indented}\item
\begin{tabular}{@{}lccc} \br
$E_{th}, eV$   &       $10^{17}$ & $10^{18}$ & $10^{19}$ \\ \mr
$N_{\gamma}$    &       430       & 135       &   28 
\\ \br
\end{tabular}
\end{indented}
\end{table}

In figure~8 we present the integral energy spectra of 
photoproduced muons. These spectra correspond 
to different values of the energy and 
arrival direction of primary $\gamma$-ray.
It is seen that for vertical air showers the spectrum
is rather soft; only $\simeq$30\% of muons have 
at  the observation level  kinetic energy
exceeding 1~GeV. However, at large arrival
angles the spectrum becomes considerably  
flatter. For example,   for $\theta=45^{\circ}$ 
about  40-45\% of muons have energy larger than 1~GeV. At the same time, 
the  energy spectrum of photoproduced muons rather weakly 
depends on the primary energy.

In order to study the relation between the numbers of 
electrons and muons  
we present in figure~9  the energy dependence of
the ratio $N_{\mu}/N_e$ for different arrival
directions. It is seen   
that for vertical showers the total number of muons 
at the observation level is very small compared with  
the number of electrons. Both $N_e$ and $N_{\mu}$ 
decrease with $\theta$, but $N_e$ decreases  
faster.  Thus at large zenith angles ($\theta\ge 65^{\circ}$) 
the number of muons   becomes comparable 
or even can exceed  the number of electrons.

\subsection{Comparison of muon content for 
air showers of different nature}
In figure~10 we present the energy dependence of  the  
total number $N_{\mu}$ of muons for vertical 
showers initiated by primary $\gamma$-rays and protons. 
The results of the present study are 
compared here with results for primary $\gamma$-rays 
obtained in Ref. \cite{Aharonian3} using  an approximate 
analytical  solution of cascade  equations,  and with the results of 
Ref. \cite{McComb,Dedenko}  for a primary proton obtained  
by the Monte Carlo  method.
On the basis of comparison of data of  Ref. \cite{Aharonian3} 
and Ref. \cite{McComb}
a conclusion has been made  in Ref. \cite{Aharonian3}
that for EHE air showers the quantity 
$N_{\mu}$ for $\gamma$-primary can be comparable with the 
number of muons in proton air showers. 
It is seen from figure~10  that  the results of present calculations  
on $N_{\mu}$ for  $\gamma$-primary (curve 1) are by a factor of a few
smaller than analytical results of 
\cite{Aharonian3} (curve 2).
This discrepancy is explained by usage in Ref. \cite{Aharonian3}
of a  simplified model of the 
photoproduction process. It was assumed that   
the mean number of charged pions created in one $\gamma$-nuclear
interaction  did not depend on the $\gamma$-ray  energy 
($E_{\gamma}$) and was equal to 2.  Furthermore,  
the energy of each of these 
pions was  assumed to be uniformly distributed  in the interval
(0,$E_{\gamma}$).  This  implies a very hard spectrum of secondary muons
and an effective  muon multiplicity of about 2.    
In fact,  the multiplicity of the muons with the total energy more than 
0.3 GeV \footnote{Exactly this value of the muon 
threshold energy is considered in figure~10.} 
approaches to 2 only for parent $\gamma$-rays with energy 
equal to several GeV. It is demonstrated in figure~12 obtained 
using the SOPHIA generator \cite{Muecke}.  On the other hand,  
the bulk of muons are produced in the electromagnetic air shower 
by cascade photons with typical energy 
of about 1 GeV for  which the muon multiplicity is 
significantly less, $\bar{n}_{\mu}\simeq 0.5$ 
(see figure~12).     
At the same time, for determination of the ratio of muon contents in 
$\gamma$- and proton induced showers,  Aharonian et al. \cite{Aharonian3}
used the calculations of  McComb et al.   \cite{McComb}.  As it is seen from 
figure~10,  the calculations of ref.  \cite{McComb} (curve 5) 
in fact predict significantly  smaller number of muons compared with 
our  calculations (curve 3).   
Thus,  the present  study does 
not confirm the conclusion of  Ref. \cite{Aharonian3} concerning the 
very high ratio of $N_{\mu}^{(\gamma)}/N_{\mu}^{(p)}$.

In figure~13 we present the  calculation results on the 
energy dependence of the ratio $\delta=N_{\mu}^{(\gamma)}/N_{\mu}^{(p)}$. 
One can
see that  this dependence has a non-trivial form. The behaviour 
of $\delta$  can be  explained by the fact the LPM effect starts 
``to operate''  a bit   earlier than the GMF does.  
In the energy region above $10^{18}$~eV 
the ratio  $\delta$ ranges within $\simeq 0.1\div 0.2$. In the energy region 
below $10^{18}$~eV it decreases  rapidly with reduction of 
the primary energy in accordance with the prediction of  
Ref.  \cite{Aharonian3}.

\subsection{Electron component}
 In figure~13  the dependence of the number of 
 electrons on the energy of primary $\gamma$-ray is
shown  for the El Nuhuil site with use of different assumptions 
about the LPM  effect  and the geomagnetic field.
One can see that the peculiarities of 
behaviour  of $N_e$ are  similar to the energy dependence of the 
muon content   described in  section~5.1 (compare figures 6 and 13).

In figure~14 we show the  results of calculations for   
the cascade curves of electrons.  Such curves can be used for an 
approximate description of the temporal  profile of the 
fluorescent radiation of air  showers.  In the case of $\gamma$-ray
primary we show the cascade curves corresponding  to both 
Northern and Southern Hemisphere sites of the Auger observatory.  
No significant  difference between the shower profiles corresponding to    
$\gamma$-ray and proton induced showers can be seen.  The 
difference in the GMF strength  between Northern and Southern
site locations (see table~1 and figure~1) 
does not affect considerably  the shower 
profile as well.  In this regard, an effective separation of 
$\gamma$-ray and proton-induced showers on the basis of the  
shower profiles seems to be hard. 
Perhaps such a separation could be achieved exploiting 
differences in the intrinsic fluctuations  of  the fluorescent light 
intensity, because the  fluctuations in    
electromagnetic showers are smaller than   
in the  hadronic ones   (see e.g.  \cite{Plyasheshnikov2}). 
The electromagnetic showers in the ``photon bunch'' regime 
caused by the GMF effect  
lead to an additional  reduction of intrinsic fluctuations and, 
therefore,  may improve  the
 $\gamma$-ray separation efficiency.

\section{Lateral distribution of photoproduced muons}
No detailed studies have been done until now  on the lateral 
distribution function (LDF)  of muons in  
electromagnetic  EHE  air showers. 
Recently,   Capdevielle et al. \cite{Capdevielle}
published  LDF of muons using   the CORSIKA 
code and taking into account the LPM effect. However, they 
ignored the effect of interactions of $\gamma$-rays with GMF,
which significantly reduces the applicability of these results. 
Moreover,  the  ``thinning technique''  
used by these authors  does not provide  an acceptable  
accuracy of calculations for the LDF of muons. 
 
The  thinning technique  has been introduced by 
Hillas \cite{Hillas2}. It allows a ``complete'' 
simulation only for a high 
energy part  of the cascade which contains secondary particles 
with energy  larger than a  fixed threshold value  $E_{\rm th}$.  
For low energy interactions, i.e. in the energy region 
$\le E_{\rm th}$,  
an artificial  absorption of secondary particles  is introduced. 
It is assumed that  only one  secondary particle survives in 
each interaction. For compensation  of this   
"disappearance"  of particles one should introduce appropriate
statistical  weights. To avoid large 
fluctuations of these  weights (dramatically  worsening  the 
accuracy of  calculations),   it is necessary 
to keep the thinning parameter 
$\epsilon = E_{th}/E$  to be
sufficiently small  ($E$ is the energy of the primary particle) . 
The computational time needed for simulation of 
one air shower by this technique  is approximately 
proportional to $\epsilon^{-1}$.

The simulations  of \cite{Capdevielle} include 
the LDF  of photoproduced muons for 10  individual 
$\gamma$-ray induced air  showers with primary energy $10^{20}$~eV. 
These results 
correspond  to the value of thinning  parameter $\epsilon=10^{-6}$. For such 
value of $\epsilon$  one needs approximately 
one hour of the computational time of 500~MHz IBM PC
to simulate one air shower \cite{Heck}.
It is seen from figure~7 of Ref. \cite{Capdevielle}  that fluctuations 
caused by introduction of statistical weights are extremely large 
(up to a factor of 30).  Therefore to  provide an acceptably small  
statistical error  (for example, 10\%) one has to 
simulate an extremely large  (up to $\simeq 10^5$)  number of
air showers. Correspondingly,  this requires a great deal of the   
computational time.   For comparison, the calculation  of LDF of 
photoproduced muons   based on the numerical solution of 
adjoint cascade equations is not laborious and requires
less than 1 hour  of the computational time.  Besides, 
this method gives  a set of LDF  in an arbitrary  energy  region  
of primaries just for {\it one run} of calculations.  
Note also that this method does not contain, by definition,   
statistical error at all. 
 
In figures~15, 16   we present the normalized LDF of   
muons of all energies, i.e.  with kinetic energy $T \geq T_{th}=0$. 
We define this parameter  as $f_{\mu}(r)=\rho_{\mu}(r)/N_{\mu}$,
where $\rho_{\mu}(r)$ is  the density of the muon flux
at distance $r$ from the shower axis. 
Results of figure~15   correspond  to different assumptions  about the GMF 
and the  LPM  effects, whereas figure~16  illustrates dependence of the 
LDF shape on the primary energy. It is seen that the geomagnetic field 
only slightly  modifies  the LDF shape.  The interaction of
$\gamma$-rays with the 
GMF makes the   LDF shape insensitive to the LPM effect.  
Also, the LDF  only slightly depends on the primary energy (see figure~16).
For example, an increase 
of $E_{\gamma}$ by a factor of 10 changes the LDF no more than 
$\simeq$20\%.
On the contrary,  at the absence of  GMF the LPM effect 
would steepen  considerably the shape of  LDF, and make it strongly 
dependent on the primary energy (figure~15).  

In figure~17  we present the muon density for vertical
proton and $\gamma$-ray induced showers. We compare in 
this  figure our results 
for primary $\gamma$-ray with the results for primary  protons 
obtained in  Ref.  \cite{Nagano}  by simulations 
with CORSIKA. One can see that the  $\gamma$-induced  
shower provides an essentially narrower 
lateral distribution of muons  compared with  the proton shower.
As a result, the ratio $\Delta=\rho_{\mu}^{(\gamma)}/\rho_{\mu}^{(p)}$
decreases rapidly with a growth of the radius
(from $\Delta\sim 0.1$ at $r=200$~m to $\Delta\sim 0.01$ at 
$r=2500$~m). This feature of $\Delta$  can  provide 
an effective separation of $\gamma$-induced air showers by 
the Auger Observatory,  for which the distance from the core 
of the  shower to the most  of triggered detectors 
is very large, several hundred meters or more. 

In figure~18  we illustrate dependence of muon LDF
on the arrival direction of the shower\footnote{
For the inclined air showers we consider the LDF averaged over
the azimuth angle in a plane perpendicular to the shower
arrival direction.}. One can see 
that the width of the LDF increases rapidly 
with a growth of the zenith angle $\theta$. This feature of LDF 
provides a significant contribution of the   
photoproduced muons to the total 
number of charged particles at large distances from 
the shower core.  This effect of inclined showers is demonstrated  in 
 figure~19, where we present the particle density  for two kinds
 of secondaries  - muons and electrons\footnote{
For the LDF of electrons we use data from Ref. \cite{Lagutin}.}. 
In our case the total number of muons
 at the observation level is much smaller than the total number of
 electrons ($N_{\mu}\simeq 0.03 N_e$; see figure~9). 
 At the same time , at distances   $r\ge 1000$~m  
from the core  of the shower  inclined at $65^{\circ}$,  
the muon density $\rho_{\mu}$ exceeds  considerably the electron 
density $\rho_e$.  This effect may have a non-negligible impact 
on  the estimates of the primary energy of inclined electromagnetic showers  
detected by air shower arrays not identifying the origin of 
secondary charged particles.  At the same time, for the 
Auger Observatory allowing  identification of secondary particles,
the detection of  unusually high $\mu/e$  ratio  in inclined showers
at large distances from the shower core should not be {\it a priori}
accepted as hadronic showers.  

\section{Conclusion}
The technique of adjoint cascade equations has been applied to study 
the characteristics of air showers created by EHE $\gamma$-rays such as 
the total numbers of cascade electrons and photoproduced muons 
at the  observation level, the longitudinal and 
lateral distributions  of  electrons and muons. Both the LPM effect and 
interactions of primary $\gamma$-rays with the geomagnetic field 
have been incorporated in the computational code.

The LPM  effect changes dramatically  
the energy dependence of the muon content $N_{\mu}$.  
Due to this effect the  growth of  $N_{\mu}$  with energy  
gradually disappears producing  a local maximum  
around  $E_{\gamma}\sim 
10^{20}-10^{21}$~eV, after which $N_{\mu}$ decreases with $E_{\gamma}$.
The interactions of $\gamma$-rays with  GMF which start to ``operate'' 
almost simultaneously with the LPM effect (i.e. at 
$E\geq 10^{19} \, \rm eV$)  recover the nominal energy dependence
of $N_{\mu}$  ``switching off'' the LPM effect.

The ratio of muon contents in the $\gamma$-ray and proton 
induced showers ranges  between $0.1-0.2$ in the EHE energy
region, but
decreases significantly at lower energies. The energy dependence 
of this quantity has a rather complicated (non power-law) form. 

Due to the combination of LPM and GMF effects the 
cascade curves of electrons in $\gamma$-ray induced showers 
are quite similar to these curves for the 
proton-induced showers.  This makes unfortunately 
rather difficult the separation 
of $\gamma$-induced air showers on the basis of the 
temporal  profiles of the fluorescent light by the 
Fly's Eye type detectors,  except a  narrow primary energy band  
(around $10^{19} \, \rm eV$) and a specific region of 
arrival directions (depending on the detector location site)
for which $\gamma$-rays freely  pass through the Earth's magnetosphere.   

The  effect of  LPM,  when combined with interactions of $\gamma$-rays
with the GMF,    on the lateral distribution of  photoproduced muons is quite 
small.  The  $\gamma$-induced air 
shower  has an  essentially narrower 
lateral distribution of muons  compared with  the proton shower.
This feature  can be effectively  used  for separation  of 
$\gamma$-ray induced air showers  by the  particle 
detectors of the  Auger Observatory.
The extension  of the  
muon LDF  increases  considerably   with the
zenith angle of  arriving primary $\gamma$-rays.
This results in an interesting effect -  the 
density of muons  at large distances from the shower core 
($\ge 1000$~m)  may noticeably  exceed  the electron  density.
 
\section*{Acknowledgements}
We are grateful to V.Sahakian for valuable
discussions.

\newpage\clearpage
\begin{figure}[p]
\centering
\includegraphics[width=0.85\textwidth]{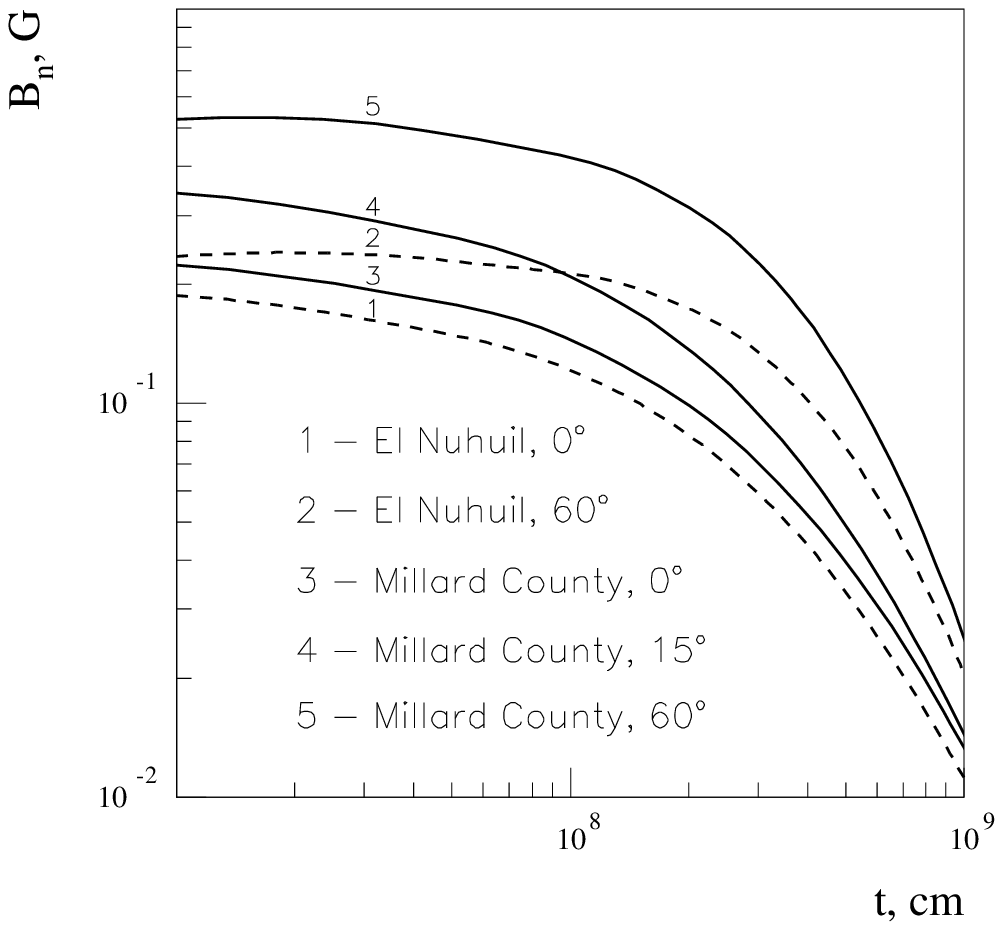}
\parbox[t]{0.9\textwidth}{ 
\caption{
Profiles of the perpendicular component of geomagnetic field
for different detector locations  and  for different zenith angles  
of arriving   $\gamma$-ray photon.}} 
\end{figure}

\newpage\clearpage
\begin{figure}[p]
\centering
\includegraphics[width=0.85\textwidth]{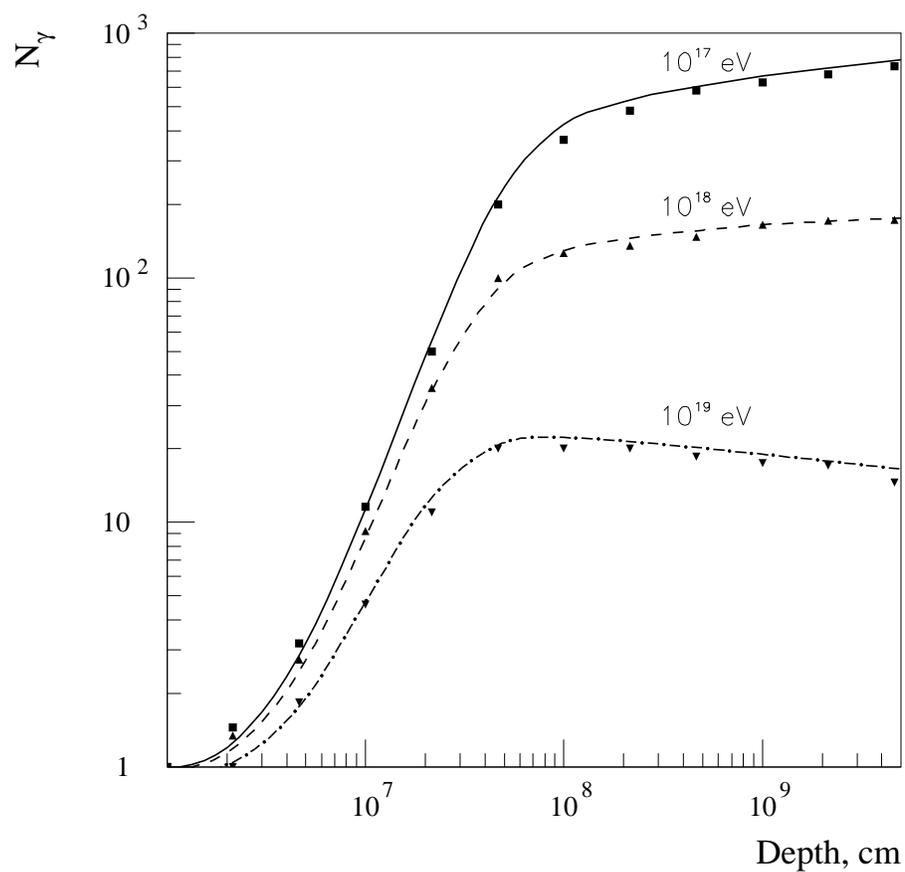}
\parbox[t]{0.9\textwidth}{ 
\caption{The number of secondary synchrotron photons with 
energies $E \geq 10^{17}, \ 10^{18}$ and  $10^{19} \, \rm eV$
produced in the homogeneous magnetic field with $B_{\rm n}=0.3 \ \rm
G$ by primary $\gamma$-rays of energy $10^{21}$~eV. 
Curves -- this work;  points  -- from Ref.  \cite{Goncharov}.}}
\end{figure}

\newpage\clearpage
\begin{figure}[p]
\centering
\includegraphics[width=0.85\textwidth]{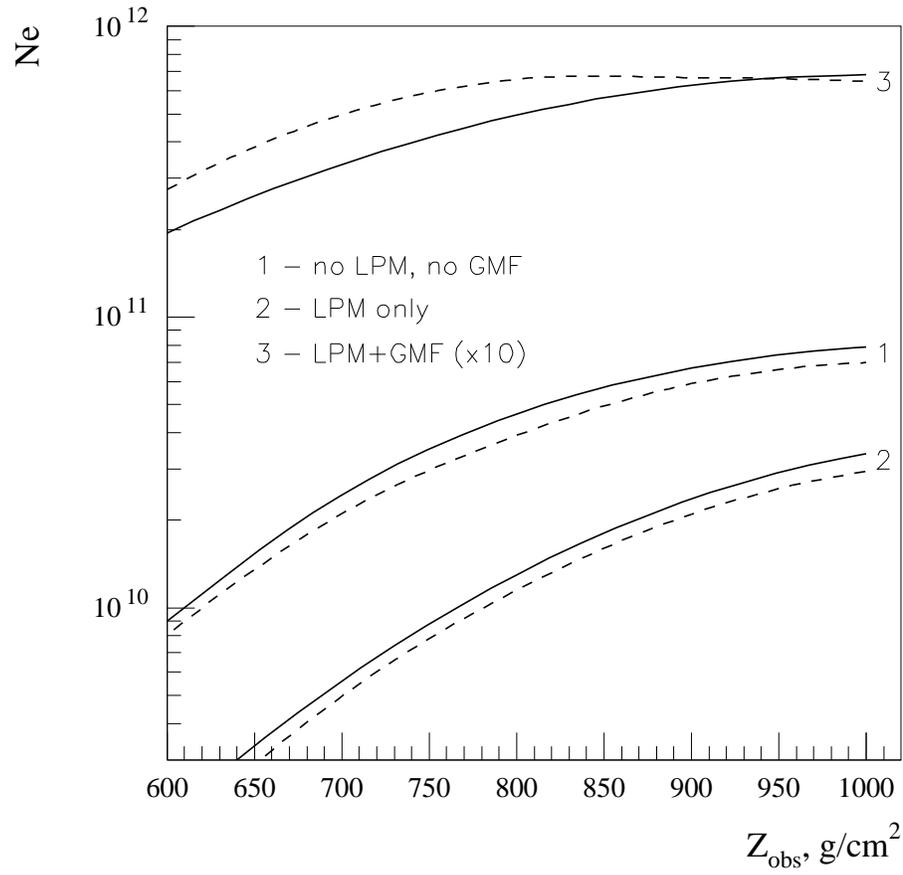}
\parbox[t]{0.9\textwidth}{ 
\caption{Dependence of the number of cascade electrons with
energy $T\ge 0$ on the  
depth of the observation  level for vertical primary  
$\gamma$-rays of  energy $10^{20}$eV with use of 
different assumptions  concerning  the LPM and GMF effects.
Solid curves  -- this work;  dashed  curves -- from Ref. \cite{Aharonian3}.}}
\end{figure}

\newpage\clearpage
\begin{figure}[p]
\centering
\includegraphics[width=0.85\textwidth]{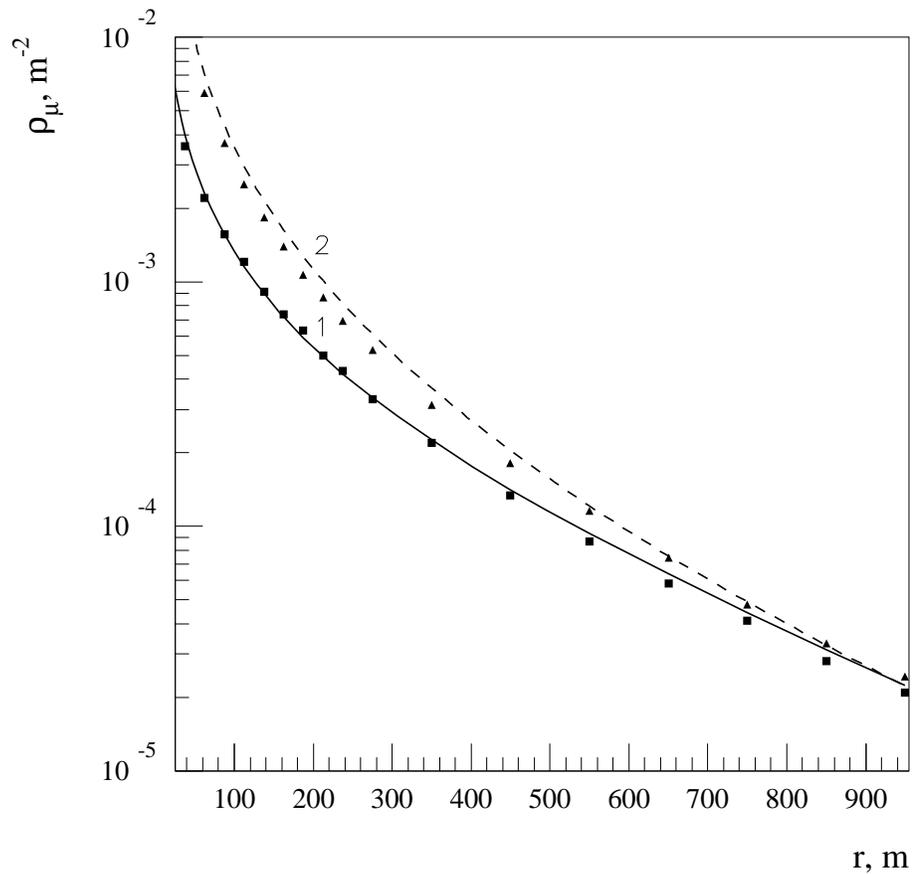}
\parbox[t]{0.9\textwidth}{ 
\caption{The radial dependence of the density of muons  with energy 
$T\ge 0$  for vertical air showers produced by primary $\gamma$-rays
with energy $E=10^{15}$~eV at the sea level (1) and at 5~km above
the sea level (2). Curves -- numerical solution of adjoint equations;
points -- simulations with CORSIKA.}}
\end{figure}

\newpage\clearpage
\begin{figure}[p]
\centering
\includegraphics[width=0.85\textwidth]{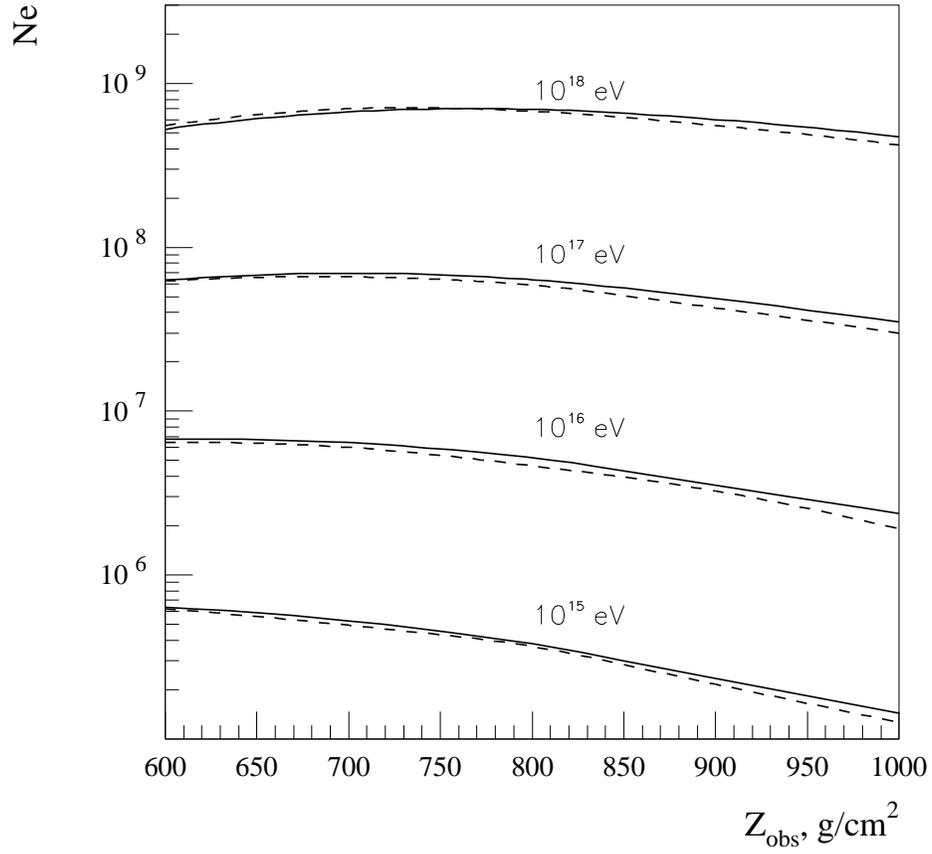}
\parbox[t]{0.9\textwidth}{ 
\caption{The number of electrons  with energy $T\ge 0$ in the  air 
showers initiated by vertical protons as a function of the penetration depth.
The energies of primary protons are indicated at the curves. 
Solid curves -- present work; 
dashed curves -- from Ref. \protect\cite{Lagutin}.}}
\end{figure}

\newpage\clearpage
\begin{figure}[p]
\centering
\includegraphics[width=0.60\textwidth]{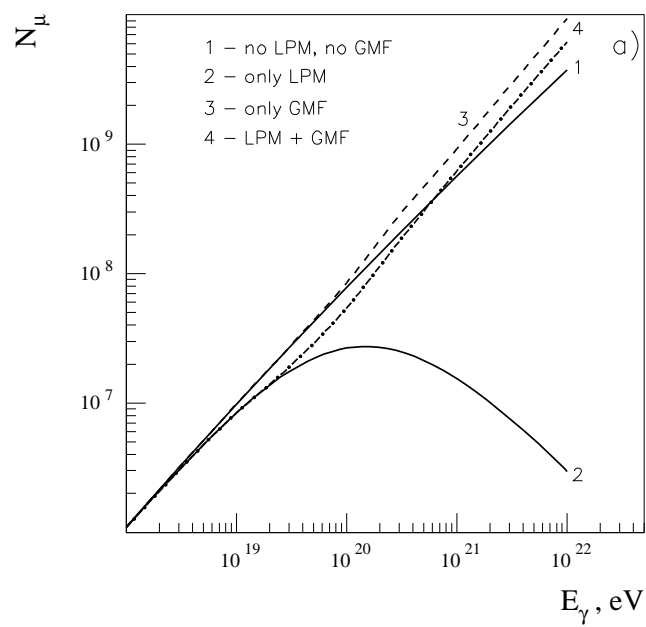}
\includegraphics[width=0.60\textwidth]{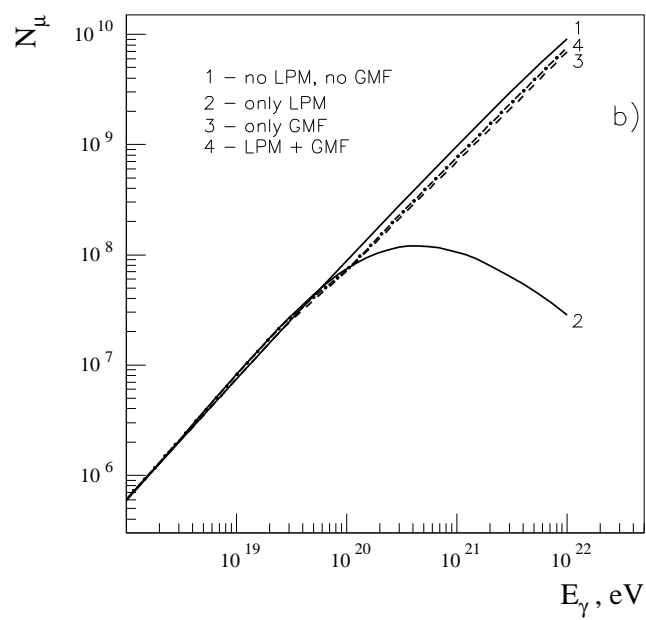}
\parbox[t]{0.9\textwidth}{
\caption{Dependence of the number of muons with energy $T\ge 0$ 
on the  energy of primary $\gamma$-ray for different assumptions 
concerning  the LPM and GMF effects calculated for the  El Nuhuil site.
Vertical (a) and  inclined ( b, $45^{\circ}$) showers are considered. }}
\end{figure}

\newpage\clearpage
\begin{figure}[p]
\centering
\includegraphics[width=0.85\textwidth]{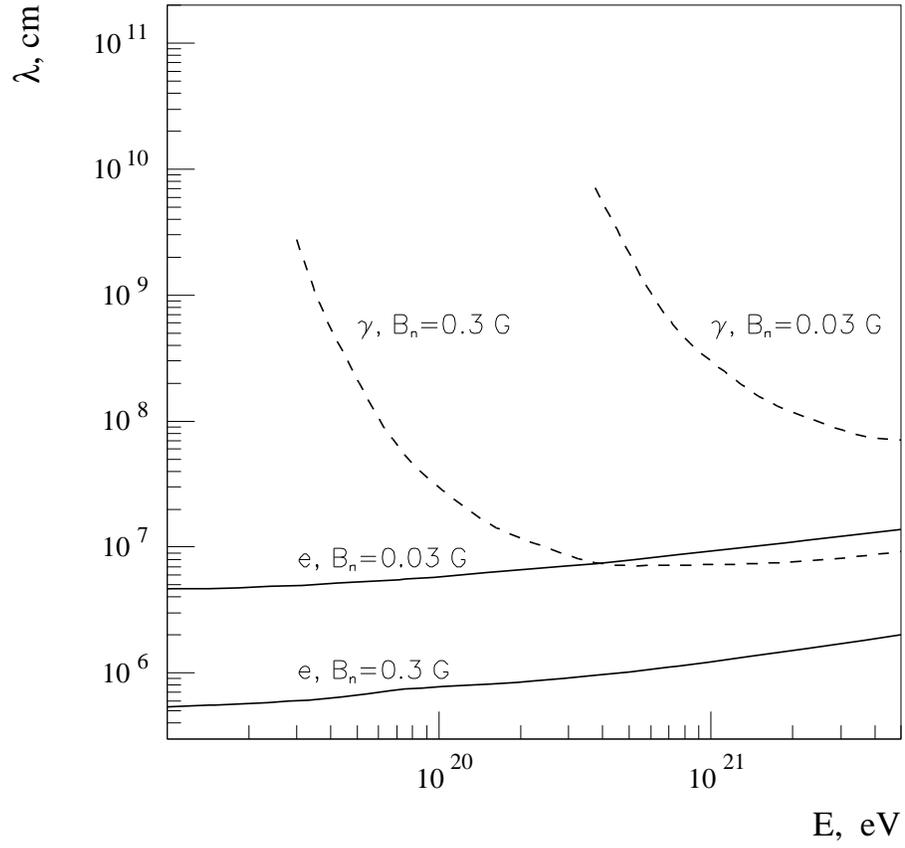}
\parbox[t]{0.9\textwidth}{ 
\caption{The energy dependence of the mean free 
pathlength of $\gamma$-rays and electrons 
in the homogeneous magnetic field $B_{\rm n}=0.3$ 
and 0.03~G.}}
\end{figure}

\newpage\clearpage
\begin{figure}[p]
\centering
\includegraphics[width=0.85\textwidth]{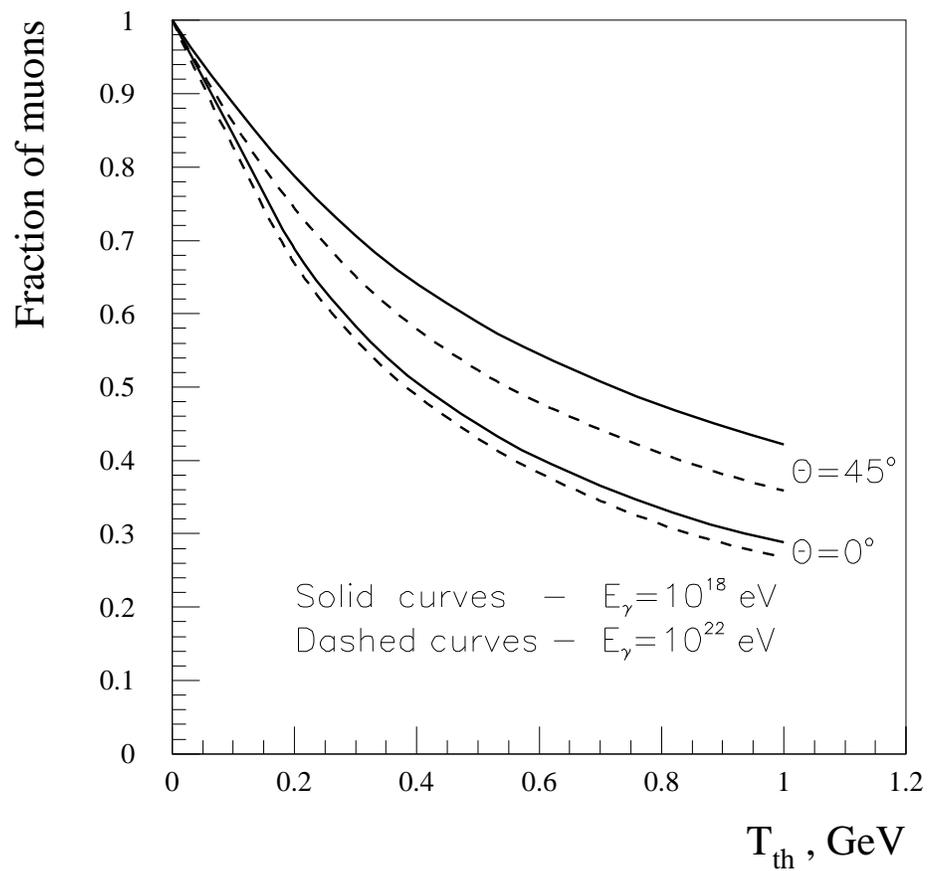}
\parbox[t]{0.9\textwidth}{ 
\caption{The fraction  of muons with kinetic
energy larger than $T_{th}$  for different incident angles 
$\theta$ and energies $E_{\gamma}$  of  
primary $\gamma$-ray.
 The El Nuhuil site.}}
\end{figure}

\newpage\clearpage
\begin{figure}[p]
\centering
\includegraphics[width=0.85\textwidth]{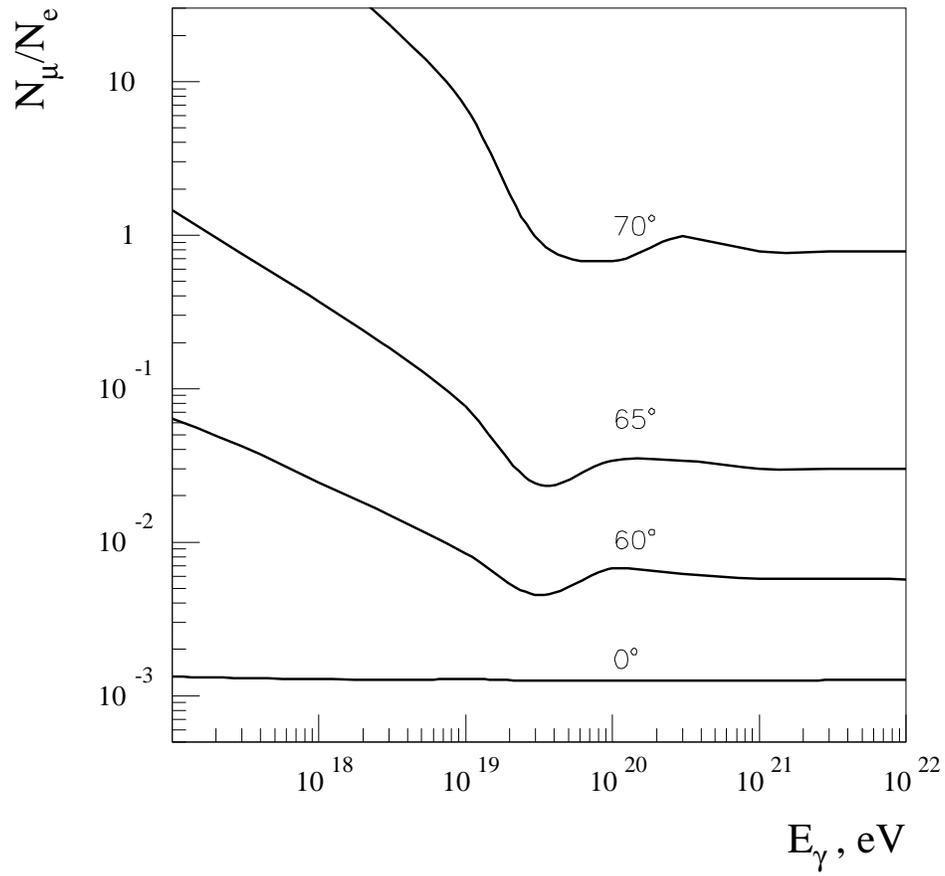}
\parbox[t]{0.9\textwidth}{ 
\caption{The ratio of total numbers of muons and electrons as a function
of energy of primary $\gamma$-ray.  The zenith angles of the arriving 
$\gamma$-rays are indicated at the curves. 
The El Nuhuil site.}}
\end{figure}

\newpage\clearpage
\begin{figure}[p]
\centering
\includegraphics[width=0.85\textwidth]{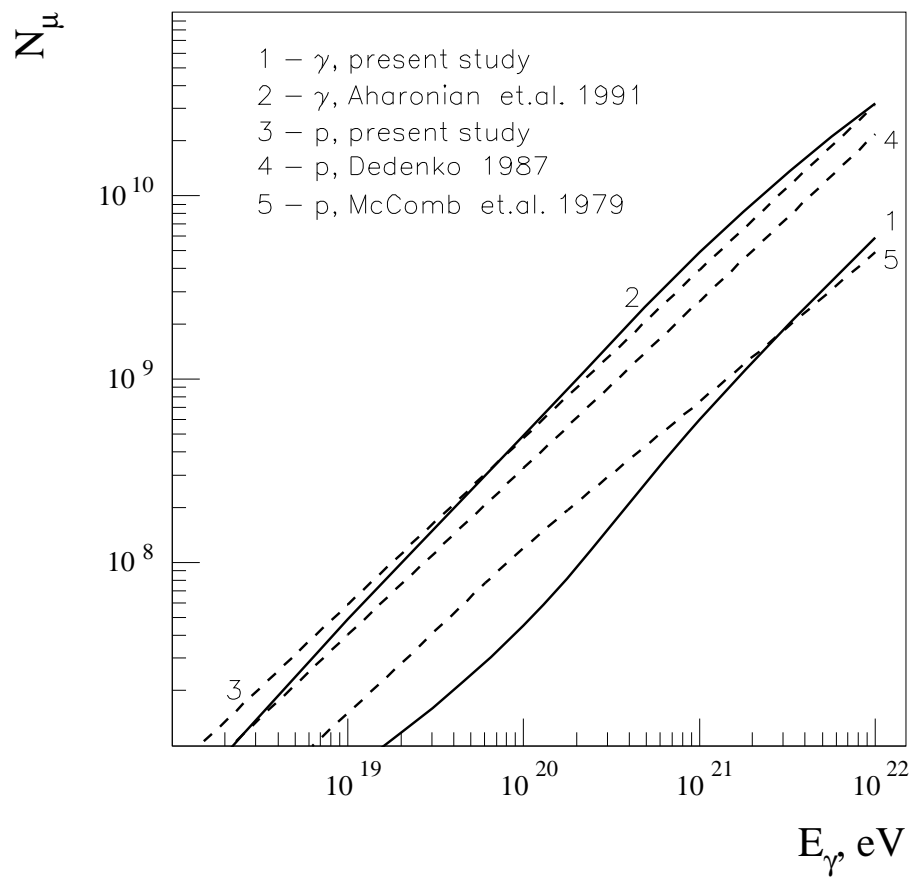}
\parbox[t]{0.9\textwidth}{  
\caption{The energy dependence of number of muons  
with total energy $E_{th}\ge 0.3$~GeV for vertical $\gamma$-ray 
and proton  induced showers at the  sea level.}}
\end{figure}

\newpage\clearpage
\begin{figure}[p]
\centering
\includegraphics[width=0.85\textwidth]{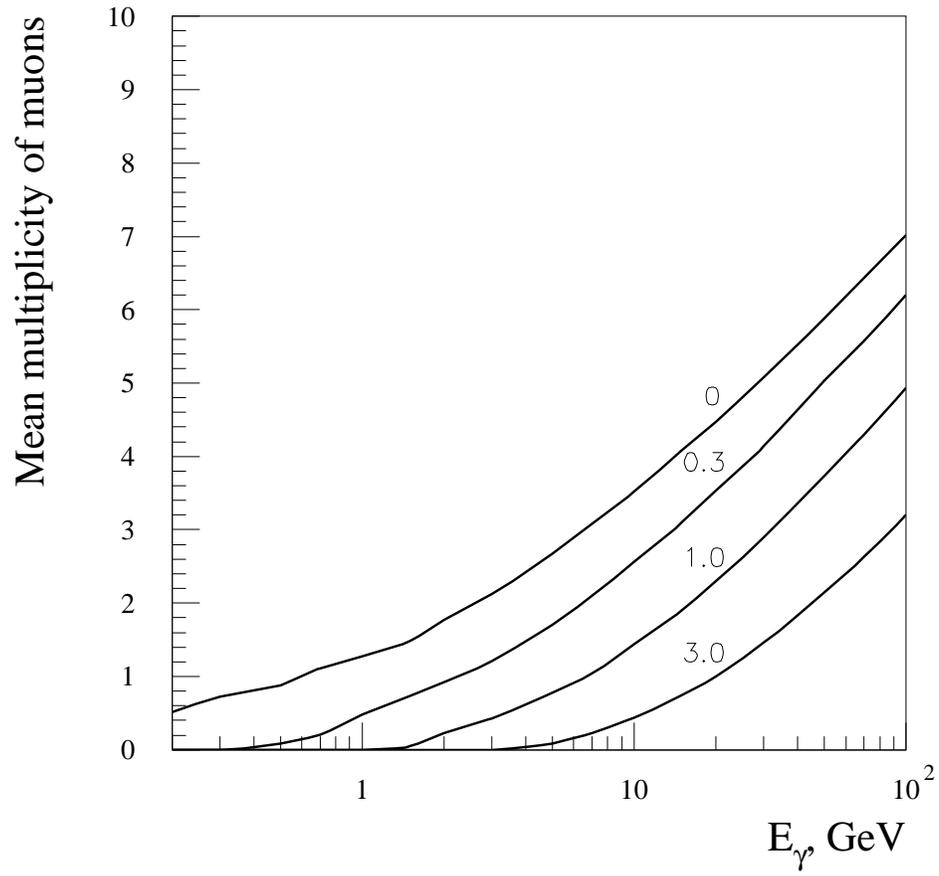}
\parbox[t]{0.9\textwidth}{ 
\caption{The mean multiplicity of muons with kinetic
energy $T\ge T_{\rm th}$
produced in the $\gamma$-nuclear interactions as a function of
the $\gamma$-ray energy. The values of $T_{\rm th}$ are indicated at
the figures.}}
\end{figure}

\newpage\clearpage
\begin{figure}[p]
\centering
\includegraphics[width=0.85\textwidth]{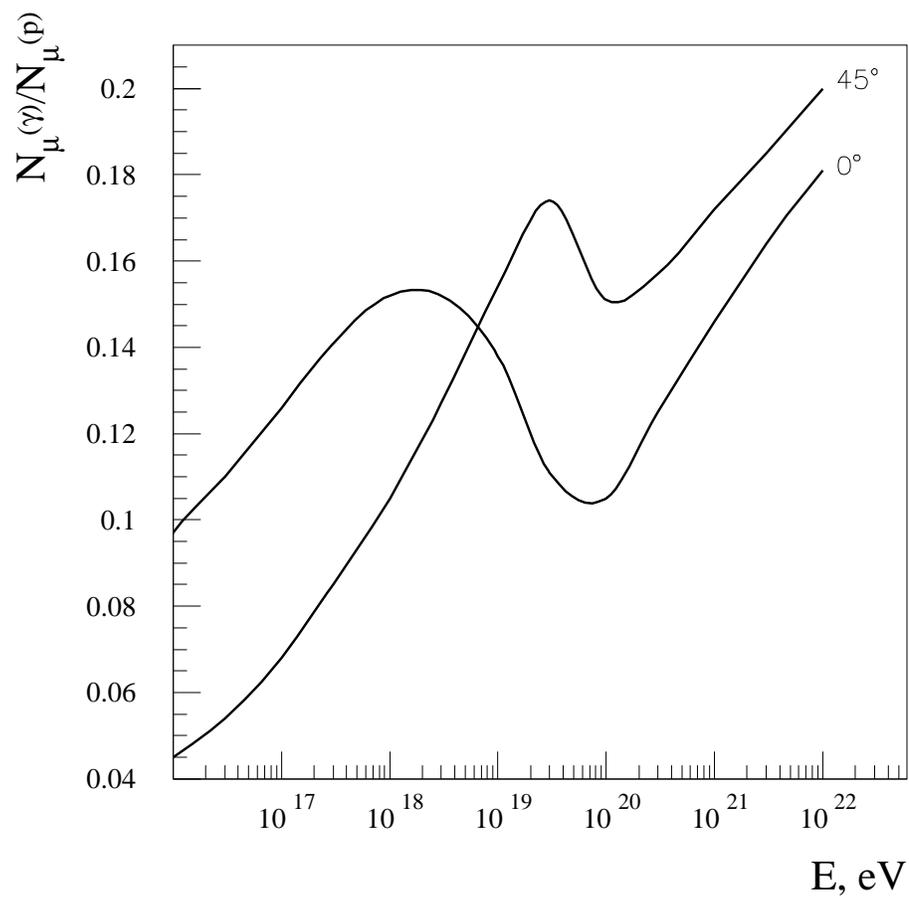}
\parbox[t]{0.9\textwidth}{ 
\caption{The ratio of the contents of muons with energy $T\ge 0$ 
for air showers produced  by primary $\gamma$-rays and protons.
The arrival  angles of primary  particles are indicated 
at the curves.  The El Nuhuil site.}}
\end{figure}

\newpage\clearpage
\begin{figure}[p]
\centering
\includegraphics[width=0.85\textwidth]{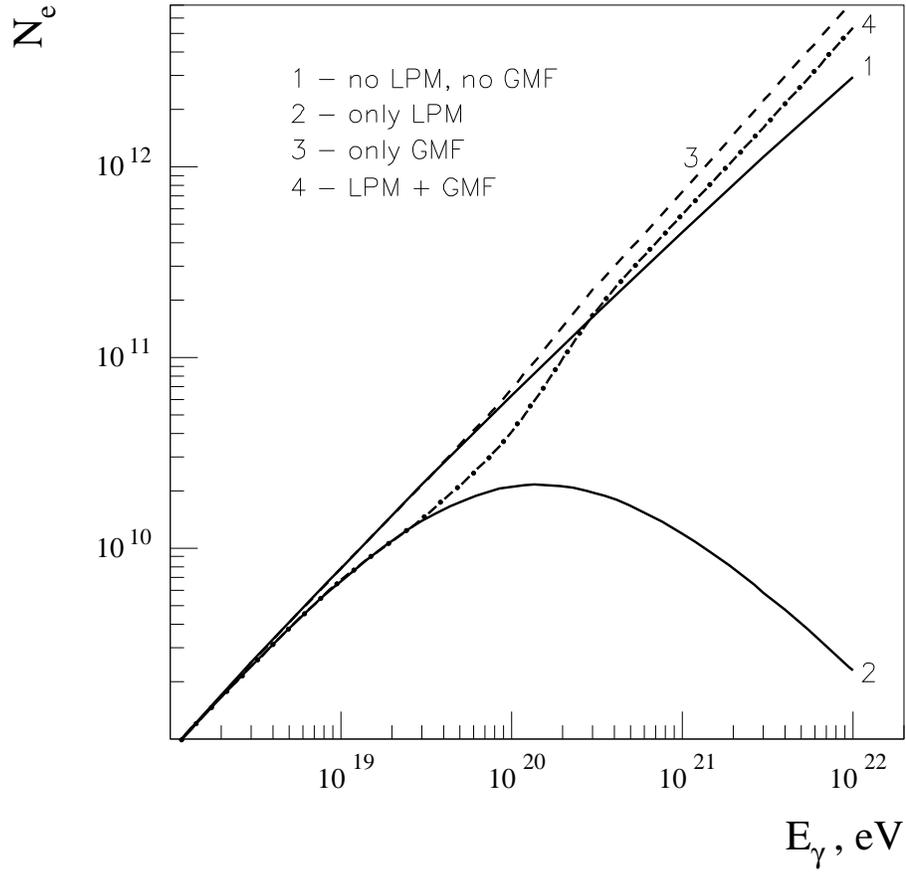}
\parbox[t]{0.9\textwidth}{ 
\caption{Dependence of the number of cascade electrons 
with energy $T\ge 0$ on the  energy of vertical $\gamma$-ray
and proton induced air showers for different assumptions 
concerning the LPM and GMF effects.  The El Nuhuil site.}}
\end{figure}

\newpage\clearpage
\begin{figure}[p]
\centering
\includegraphics[width=0.85\textwidth]{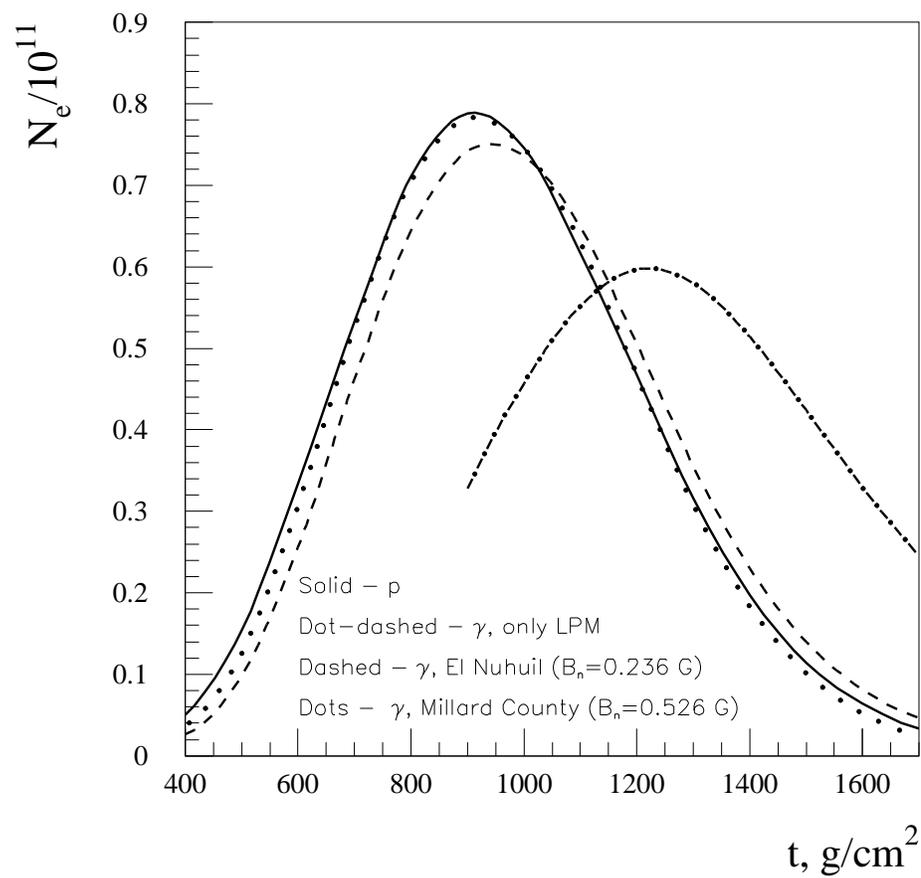}
\parbox[t]{0.9\textwidth}{  
\caption{The number of cascade electrons with energy $T\ge 0$ 
as a function of the penetration depth
for primary protons and $\gamma$-rays 
of energy $10^{20}$~eV. Two different detector locations  -- 
El Nuhuil and  Millard County -- are considered  for  
inclined (60$^{\circ}$) air showers.}}
\end{figure}

\newpage\clearpage
\begin{figure}[p]
\centering
\includegraphics[width=0.85\textwidth]{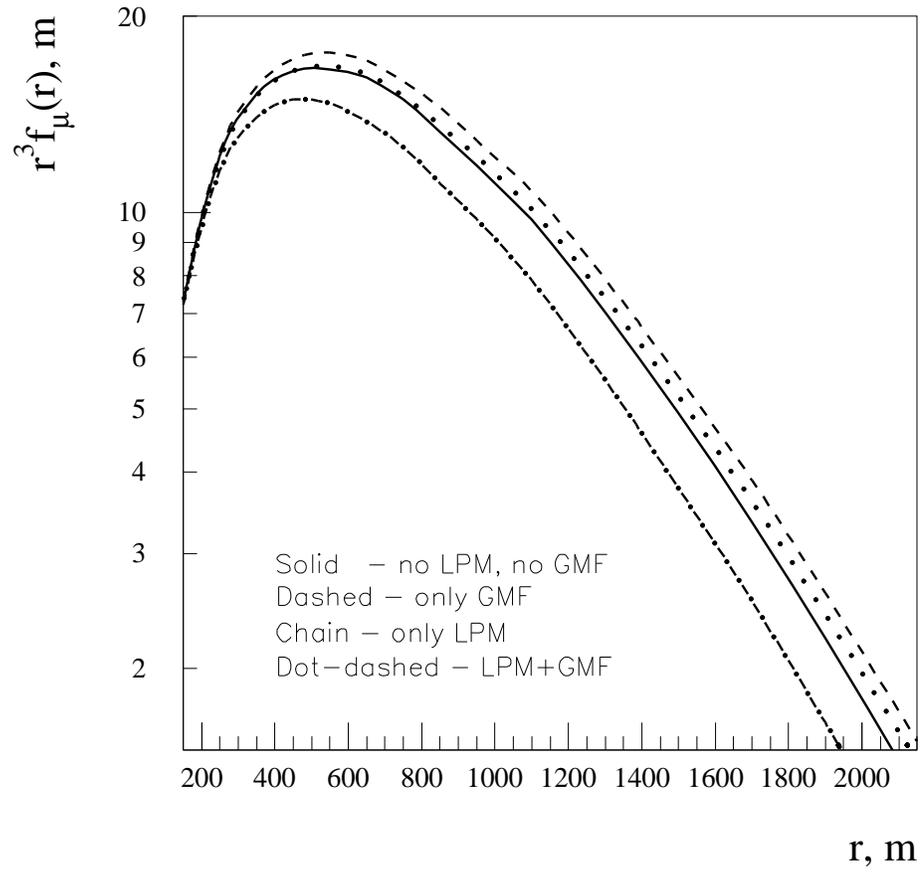}
\parbox[t]{0.9\textwidth}{ 
\caption{The lateral  distributions of muons 
with energy $T\ge 0$ for vertical showers induced by 
a primary $\gamma$-ray of energy  $10^{20}$~eV, 
calculated for different assumptions 
concerning the LPM and GMF effects.  The El Nuhuil site.}}
\end{figure}

\newpage\clearpage
\begin{figure}[p]
\centering
\includegraphics[width=0.85\textwidth]{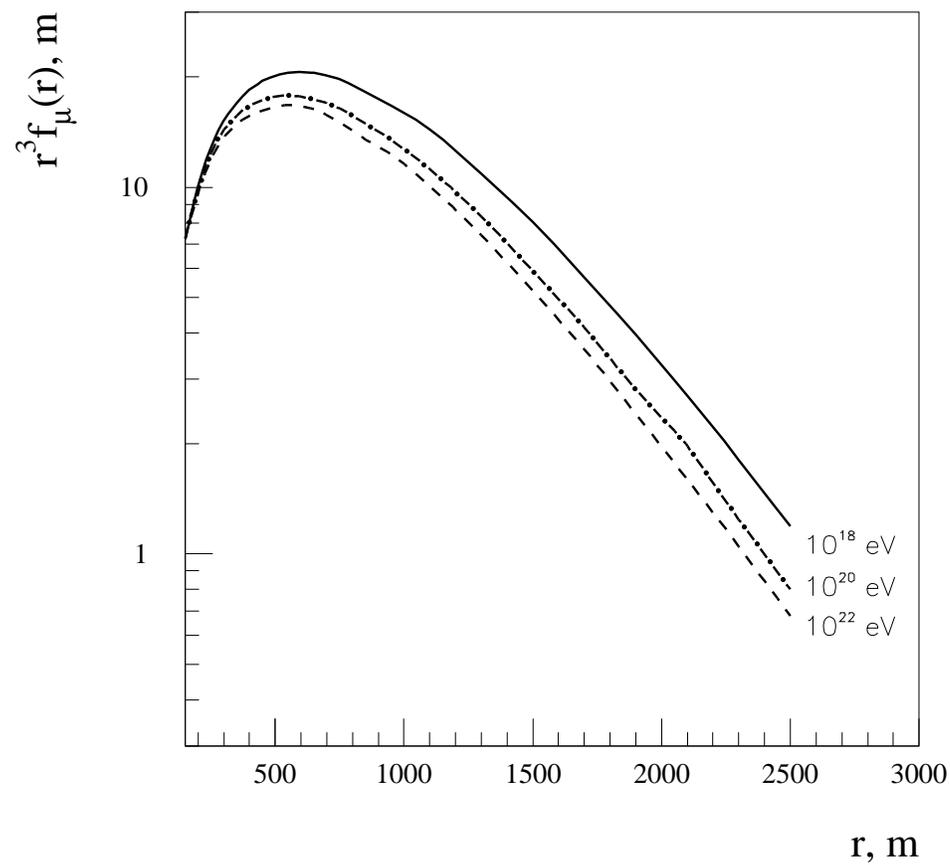}
\parbox[t]{0.9\textwidth}{  
\caption{The lateral distributions of muons with energy $T\ge 0$  
for  vertical  $\gamma$-ray showers of energies indicated at the curves.
The El Nuhuil site.}}
\end{figure}

\newpage\clearpage
\begin{figure}[p]
\centering
\includegraphics[width=0.85\textwidth]{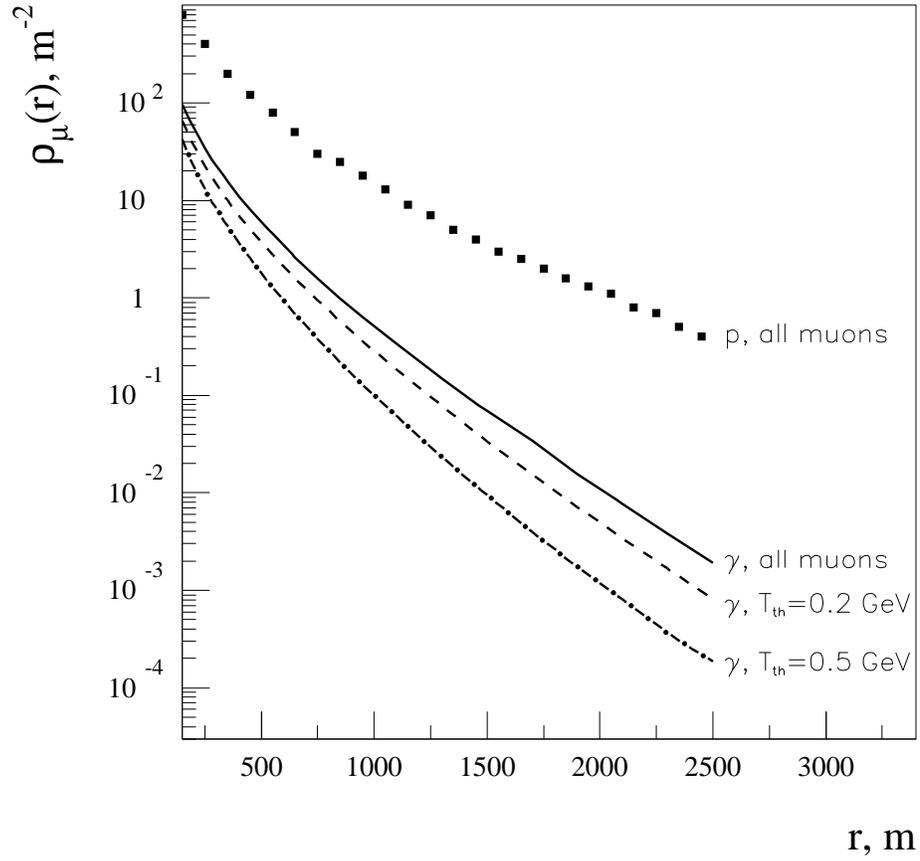}
\parbox[t]{0.9\textwidth}{ 
\caption{The muon density in the  vertical
showers produced by protons and $\gamma$-rays of energy  
$10^{20}$~eV. The results of present work for primary $\gamma$-ray 
(the El Nuhuil site) and for three different muon threshold energies 
are compared with results of Ref.  \cite{Nagano}
calculated  for primary protons  at  $Z_{\rm obs}=920$ g/cm$^2$  .}}
\end{figure}

\newpage\clearpage
\begin{figure}[p]
\centering
\includegraphics[width=0.85\textwidth]{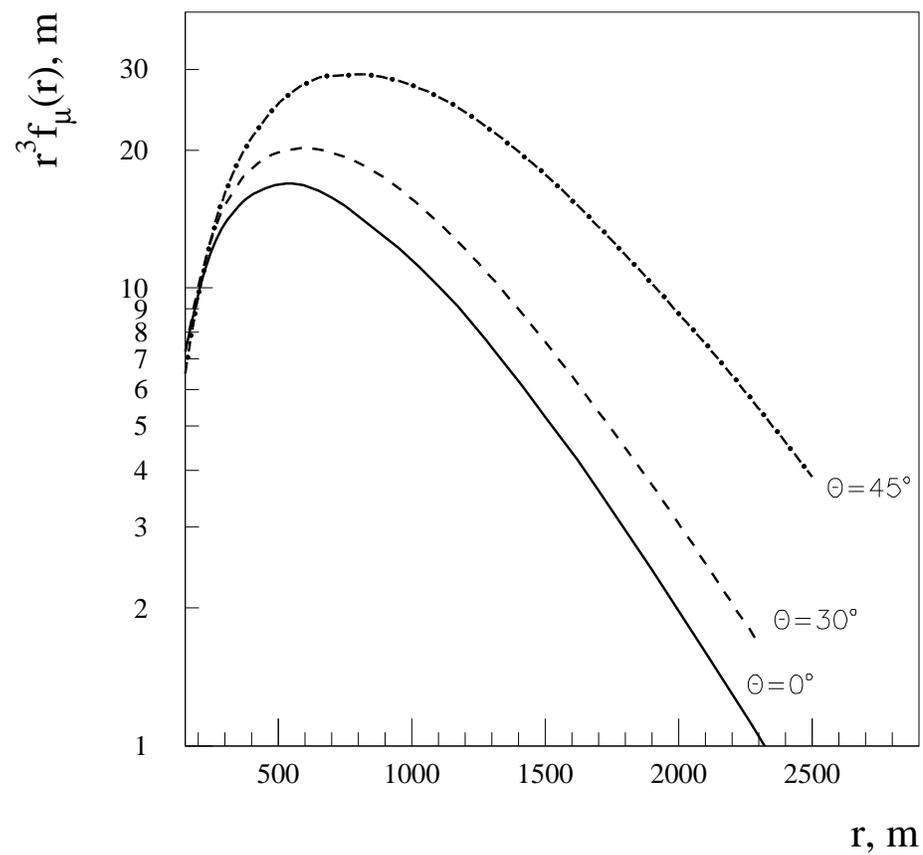}
\parbox[t]{0.9\textwidth}{ 
\caption{The lateral distributions  of  muons 
with energy $T\ge 0$ in the electromagnetic showers 
for 3 different arrival angles of primary $\gamma$-ray of  energy 
$10^{20}$~eV.  The El Nuhuil site.}}
\end{figure}

\newpage\clearpage
\begin{figure}[p]
\centering
\includegraphics[width=0.85\textwidth]{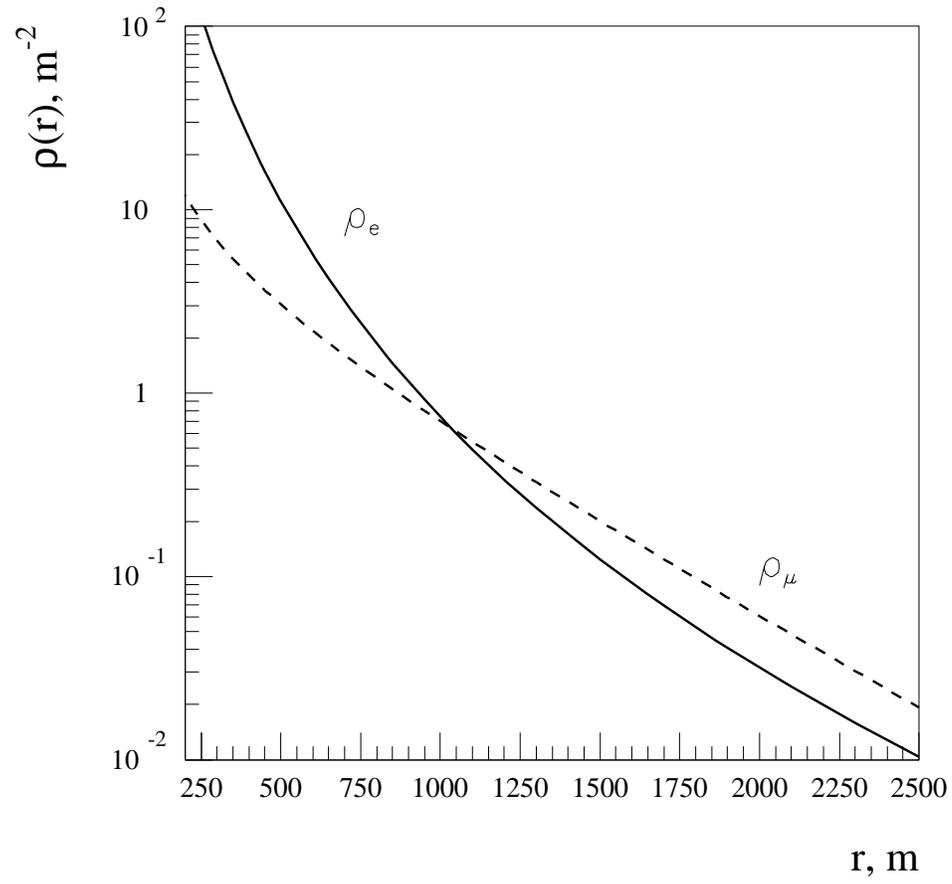}
\parbox[t]{0.9\textwidth}{ 
\caption{The densities of muons  ($\rho_{\mu}$) and  
electrons  ($\rho_e$) in the inclined ($65^{\circ}$)
$\gamma$-ray shower of energy $10^{20}$~eV.  The El Nuhuil site.}}
\end{figure}

\end{document}